%% file: YangNJPSep27.tex
\documentclass[]{iopart}

\usepackage{iopams}

\newcommand{\hc}{H_{\mathbb{C}}}
\newcommand{\hq}{H_{\mathbb{H}}}
\newcommand{\rthree}{{\mathbb{R}^3}} 
\newcommand{\rfive}{{\mathbb{R}^5}} 
\newcommand{\qi}{{i_\mathbb{H}}} 
\newcommand{\qj}{{j_\mathbb{H}}} 
\newcommand{\qk}{{k_\mathbb{H}}} 

\usepackage[]{graphicx}
\usepackage[]{color}
\usepackage[]{amssymb}
\usepackage[]{amsfonts}
\usepackage[]{latexsym}
\usepackage[]{float}
\usepackage{bm}

\input{mydef.tex}
\newcommand{\mpic}[1]{}

\begin{document}

\title[
Symmetry protected $\mathbb{Z}_2$-quantization and
quaternionic Berry connection with KR degeneracy
 ]
{Symmetry protected $\mathbb{Z}_2$-quantization and quaternionic Berry connection with Kramers degeneracy}

\author{Yasuhiro Hatsugai}

\address{
Institute of Physics, 
University of Tsukuba, 1-1-1 Tennodai, Tsukuba, Ibaraki 305-8571, Japan
}
\ead{y.hatsugai@gmail.com}
\begin{abstract}
As for a generic parameter dependent hamiltonian with
the  time reversal (TR) invariance, 
a non Abelian Berry connection 
with the Kramers (KR) degeneracy
are introduced by using a quaternionic Berry connection. 
This quaternionic structure naturally extends to the 
many body system with the KR degeneracy. 
Its topological structure is explicitly discussed in 
comparison with the one without the KR degeneracy. 
Natural dimensions to have non trivial topological structures
are discussed by presenting explicit gauge fixing.
Minimum models to have accidental degeneracies are
given with/without the KR degeneracy, 
which describe the monopoles of Dirac and Yang.
We have shown that the
 Yang monopole is literally a quaternionic Dirac monopole.

The generic Berry phases with/without the KR degeneracy are
introduced by the complex/quaternionic Berry connections.
As for the symmetry protected
 $\mathbb{Z}_2$ quantization of these general Berry phases,
a sufficient condition of the $\mathbb{Z}_2$-quantization 
is given as the inversion/reflection equivalence.

Topological charges  of the $SO(3)$ and $SO(5)$
 nonlinear
$\sigma $-models 
are
discussed in their relation to the Chern numbers 
of the 
$CP^1 $ and $HP^1$ models as well.
\end{abstract}

\maketitle

\section{Introduction}
\label{sec:intro}

Topological numbers have been important in physics 
especially in quantum phenomena.
They give a conceptual foundation of
quantizations for various elementary degrees of freedom such as
 charges, fluxes, vortices and monopoles\cite{dirac,Thouless98}.
One of the milestones of the emerging topological numbers
is a quantization of the Hall conductance
where a response function is directly related to the topological 
quantum number as the first Chern number
\cite{Thouless82,Simon83,NTW,Kohmoto85,Hatsugai93b}. 
Its fundamental physical  meaning has become clear by introducing
an idea of the geometrical concept which is known as the  Berry connection
today\cite{Berry84}.
For the quantum Hall (QH) states, 
the bulk is gapped and does not have any characteristic symmetry breaking.
It results in absent of a local order parameter nor
any low energy excitations as the Goldston bosons.
A class of such featureless systems is the  (gapped) 
quantum liquid and the spin liquid.
A possible effective field theory for the gapped quantum liquids
is the topological field theory where topological quantities
play a central role. Then corresponding new idea to
describe the system is the topological order\cite{Wen89,wen_topological2}.
It should be compared with the standard Ginzburg-Landau-Wilson scenario,
where local field theories to describe fluctuation of a
local order parameter is essential. 
As for a bulk topological ordered state, the degeneracy 
of the ground state depends
on the topology of the physical space\cite{Wen89}.
However there were not so much quantities to describe the topological order.
As an extension of the success for the QH state, we have successfully 
used the Berry connections and related topological quantities for
characterization of the topological ordered 
states\cite{Hatsugai04e,Hatsugai05-char,Hatsugai06a,HiranoTopo,HiranoDeg,MaruSpinBerry,RingExBerry}.
Also note that although the bulk QH state is
featureless,
the system with boundaries
has characteristic localized states 
as the edge states\cite{Halperin82,Hatsugai93a}.
Extending the observation, 
we are proposing an idea of the {\em bulk-edge correspondence}, 
which says that
although 
the bulk is gapped and only characterized by the topological quantities,
there exist characteristic boundary states
which reflect the topologically non trivial bulk 
for the system with boundaries\cite{Hatsugai93b,yh-qhe-review}.
This "bulk-edge correspondence"
 seems to be a universal 
feature of the topological ordered states
such as the QH states, 
quantum spins\cite{Hatsugai06a,KatsuEnt,RingExBerry},
graphene\cite{HatsuSSC},
photonic crystals\cite{edgePhotonic},
cold atoms\cite{edgeColdAtom}, 
characterization of localizations\cite{Schnyder08}
 and
quantum spin Hall (QSH) systems\cite{KM05,BHZ,Qi08}.

The QSH state
is an analogous state to the QH states but it respects
the time reversal symmetry by the help of spins\cite{KM05,BHZ}. 
Then it is natural that
the Berry connection
with the TR invariance play
 fundamental roles.
There have been  substantial amount of works for the topic
\cite{Avron88,Demler99,Murakami04,BHZ,KM05,FuKane06,FuKane07,
Fukui07,Fukui08,Qi08,
Schnyder08,
SurMoore,EssinMooreVan,Ryu09,Schnyder09}.
Here
in this paper, we present a self-contained description of
the Berry connections and related topological quantities
with/without the Kramers (KR) degeneracy. 
Especially we focus 
on its quaternioninc structure. 
The quaternions
are fundamental in the description of the 
TR invariant system with the KR degeneracy
which was first pointed out by 
Dyson long time ago\cite{Dyson62,Zumino62,Avron88,Demler99}.
There is more than an analogy between the system with/without the 
KR degeneracy.
One can make a mapping between them by 
replacing the complex number by the quaternions\cite{Zumino62}. 
We explicitly demonstrate it for the topological quantities
by introducing canonical minimum models, which are related 
to the monopoles and accidental degeneracies.

As for the topological quantities, there can be two classes. The
one includes  topological invariants by their definition.
The quantization for them is automatically guaranteed only by 
 stability and a regularity of the 
Berry connections. The examples are  the Chern numbers,
 winding numbers and Pontrjagin index. 
Additionally we introduce a new class of quantized quantities
as a generalization of the $\mathbb{Z}_2$ 
Berry phase\cite{Hatsugai06a},
which is geometrical but as for the quantization, 
one needs additional symmetry requirement.
We give a sufficient condition for this
symmetry protected  $\mathbb{Z}_2$-quantization.

As for the application of gauge invariant description of the
Chern numbers, a relation between the Chern numbers and 
the topological charges of the $SO(3)$ and $SO(5)$ nonlinear $\sigma$-models
are also presented shortly.


\section{ Time reversal and quaternions}  
\label{sec:trq}
Let us first introduce a quaternion notation 
for a TR invariant bi-linear system\cite{Dyson62}.
Introducing  $D$ parameters  $x=(x^1,\cdots,x^\mu,\cdots,x^D)\in V_D$, 
${\rm  dim}\, V_D=D$, 
let us consider 
a bi-linear $2N$-fermion hamiltonian 
${\cal H} (x)={c}_m ^\dagger {H}_{mn}(x) {c}_n $,
${c}_n ^\dagger  =(c_{n\uparrow} ^\dagger ,c_{n\downarrow} ^\dagger)$ 
where 
 $H_{mn}$ is a $2\times 2$ complex matrix and 
 $c_n$, ($n=1,\cdots,N$) is a spinor, a pair of fermion annihilation operators
(summation over doubled indexes is assumed and $n=1,\cdots,N$). 
Further let us impose an invariance under the time-reversal (TR) 
operation $\Theta$ for the hamiltonian ${\cal H}$.
Since $\Theta$ operates as
 $c_{n \sigma }\to(-)^{(\sigma-1)/2} c_{n-\sigma }$ (
${\uparrow} =+1$ and 
${\downarrow} =-1$, 
$c_{n {\uparrow}  }\to c_{n {\downarrow} } $ and
$c_{n {\downarrow}  }\to - c_{n {\uparrow} } $ 
) 
and taking a complex conjugate
 $\cal K$,
we have  $\widetilde {J} H_{mn}^*J =-J H_{mn}^*J 
= H_{mn}$,
($J=i \sigma _y$) where $\sigma_{x,y,z} $ are the Pauli matrices
($\, \widetilde{\ }$ is a matrix transpose).
As for the bi-linear hamiltonian here, 
it is expressed as $[H,\Theta_b]=0$ where $\{H\}_{mn}=H_{mn}$ and
 $\Theta_b=-{\cal K} J$ ($J$ operates sub block of $H_{mn}$).
Now let us expand this $2\times 2$ matrix $H_{mn}$ as
$H_{mn}=h_{mn}^0+h^1_{mn} I + h^2_{mn} J + h^3_{mn} K$ where
 $I 
=i\sigma_z=-I^*=-I ^\dagger $,
 $J 
=i\sigma_y=J^*=-J ^\dagger $,
 $K 
=i\sigma_x=-K^*=-K ^\dagger $.
Then the TR invariance implies 
$h^\alpha_{mn}
\in \mathbb{R}$, ($\alpha=0,\cdots,3$),
that is, $H_{mn}$ is identified as a quaternion $\mathbb{H}\ni
h_{mn}$ by a standard equivalence 
$I\cong \qi $, $J\cong \qj $, $K\cong \qk $,  
$\qi,\qj,\qk
\in \mathbb{H}$, 
$
\qi^2=\qj^2=\qk^2=\qi\qj\qk=-1$,
since 
$ 
-J H_{mn}^* J=
(h_{mn}^0)^*(-J J)
+(h_{mn}^1)^*( -J(-I)J )
+(h_{mn}^2)^*( -J(J)J )
+(h_{mn}^3)^*( -J(-K)J )
\cong
(h_{mn}^0)^*
+(h_{mn}^1)^* \qi
+(h_{mn}^2)^* \qj
+(h_{mn}^3)^* \qk
$.
Hermiticity of the ${\cal H}$ , $H ^\dagger  = H$, implies
4 conditions for the
{\em real} matrices, $h^\alpha   $, $\widetilde {h}^0= h^0$, 
$\widetilde {h}^\alpha  =-h^\alpha $, ($\alpha =1,2,3$)
where $(h^\alpha )_{mn}\equiv h_{mn}^\alpha $.
It gives a hermite quaternionic matrix 
$h ^{\mathbb{H}}=h^0+h^1\qi+h^1\qi+h^2\qj+h^3\qk
 \cong H=h^0+h^1 I+h^2 J + h^3 K$ expressed as  
$(h ^{\mathbb{H}})  ^\dagger   = h ^{\mathbb{H}}  $.

As for a normalized eigen state, 
$\psi_\ell=\mvec{\psi_{\ell\uparrow}}{\psi_{\ell\downarrow}} $, 
($\psi_\ell ^\dagger \psi_\ell=1$)
of $2N$ dimensional hamiltonian $H$,
 ($H\psi_\ell=\epsilon_{\ell}\psi_\ell $), 
it is degenerate with 
$\psi_\ell ^\Theta=\Theta\psi_\ell=\mvec{-\psi_{\ell\downarrow}^*}{\psi_{\ell\uparrow}^*} $,
which is the Kramers(KR) degeneracy.
Its orthogonality, $\psi_\ell ^\dagger \psi_\ell^\Theta=0$,
is trivial here 
(Generically, there are $N$ KR pairs, $\ell=1,\cdots,N$).
Then let us write this KR pair as
\begin{eqnarray*} 
\fl\qquad 
 \Psi_\ell =  (\psi_\ell,\Theta\psi_\ell)
&=  
 \psi_\ell^0 \otimes E_2
+ \psi_\ell^1 \otimes I
+ \psi_\ell^2 \otimes J
+ \psi_\ell^3 \otimes K
\\
\fl\qquad 
&= 
\mmat
{\psi_\ell^0+i\psi_\ell^1}{\psi_\ell^2+i\psi_\ell^3}
{-\psi_\ell^2+i\psi_\ell^3}{\psi_\ell^0-i\psi_\ell^1}
\cong \psi_\ell ^{\mathbb{H}} \in\mathbb{H}^N,
\end{eqnarray*} 
where 
$\psi^0_\ell={\rm Re\,} \psi_\ell^{\uparrow}  $,
$\psi^1_\ell={\rm Im\,} \psi_\ell^{\uparrow}  $,
$\psi^2_\ell=-{\rm Re\,} \psi_\ell{\downarrow}  $,
$\psi^3_\ell={\rm Im\,} \psi_\ell^{\downarrow}  $, 
$\psi_\ell^\alpha \in \mathbb{R}^N$ and
$E_2$ is a two-dimensional unit matrix.
Here 
$\psi_\ell ^{\mathbb{H}}$
is a quaternion vector of dimension $N$.

A linear canonical transformation of the fermions $\{c_n\}\to \{d_\ell\}$, 
$c_n = U_{n\ell}d_\ell $, which is consistent with the TR invariance,
(written in  $\{d_\ell\}$) requires that $2\times 2$ matrix
$U_{n\ell}$ does commute with the time
reversal, that is, 
$\tilde {J} U^*_{n\ell}J=U_{n\ell}\cong u_{n\ell} ^{\mathbb{H}} \in \mathbb{H}$.
Supplementing the unitarity of this matrix $U ^\dagger U=U U ^\dagger =E_{2N}$,
$(U)_{n\ell}=U_{n\ell}$, $U\in U(2N,\mathbb{C})$, 
which guarantees the 
fermion anticommutation relations of $\{d_{\sigma \ell}\}$'s,
this $2N\times 2N$ matrix $U$ satisfies $\widetilde {U}J_{2N}U=J_{2N} $,
$J_{2N}=J\otimes E_N$, ($U\in Sp(2N,\mathbb{C})$).
It implies $U\in Sp(N)=U(2N,\mathbb{C})\cap Sp(2N,\mathbb{C})$ as a 
$2N$-dimensional  matrix.
By the standard equivalence, we also have an $N$-dimensional 
quaternion matrix 
$u ^{\mathbb{H}} \in
 M_N(\mathbb{H})$, $(u ^{\mathbb{H}} )_{n\ell}= u_{n\ell}^{\mathbb{H}}
 \in\mathbb{H}$.
It is constructed from all of the orthonormalized 
eigen states (KR pairs), $\{\psi_\ell ^{\mathbb{H}}\} $, as 
$u ^{\mathbb{H}} =(\psi_1 ^{\mathbb{H}} ,\cdots, \psi_N ^{\mathbb{H}} )$,
$H\psi_\ell=\epsilon_{\ell}\psi_]\ell $ ($\epsilon_\ell\ne \epsilon_{\ell'},
\ell\ne\ell' $), 

\section{
Quaternionic structure of many body system with  KR degeneracy}
\label{sec:many}
The quaternionic structure introduced in the previous 
Sec.\ref{sec:trq} 
is  directly extended to the Fock space of the fermion many body states
as far as the total number of particles is conserved, 
since the TR operation $\Theta$ does commute with the
$Sp(N)$ unitary transformation among the fermion spinors
$\{\bi{c}_n ^\dagger\}\to\{\bi{d}_n ^\dagger\} $
and the TR operation $\Theta$,
$\bi{c}_{i {\uparrow} }\to \bi{c}_{i {\downarrow} }$,
$\bi{c}_{i {\downarrow} }\to -\bi{c}_{i {\uparrow} }$
and
taking the complex conjugate, has a basis independent 
meaning.
Then it is also applicable for the $S=\case 1 2 $ quantum spins by the
standard representation 
$\bi{S}_i= \case 1 2\bi{c}_i ^\dagger \bi{\sigma }\bi{c}_i $
(extension to the half-odd-integer spins is trivial
by introducing the Hund coupling). 

Now let us consider a  TR invariant many body hamiltonian ${\cal H}$,
$ [{\cal H},\Theta]=0$.
When the state $|\Psi \rangle $ is an eigen state of the hamiltonian,
its time-reversal pair 
$
|\Psi^\Theta \rangle 
=
\Theta|\Psi \rangle 
$ is also an 
eigen state. 
As commented before,
we assume the hamiltonian preserves the total fermion number.
Then
one may discuss an 
$M$ particle sector separately.
The TR operation for this $M$ particle sector is then satisfy
$
\Theta ^2|\psi \rangle = (-)^M|\psi \rangle 
$.

Now let us further 
assume that the number of total fermions ($1/2$ spins) $M$ are odd to
have the KR degeneracy. Then we have the following fundamental relation
\begin{eqnarray*}
\Theta ^2|\psi \rangle = -|\psi \rangle .
\end{eqnarray*}
A generic $M$ particle state is spanned by the Fock basis as
\begin{eqnarray*}
|\psi \rangle =& \sum
\big[
\psi_O(i)|O(i) \rangle +
\psi_E(i)|E(i) \rangle \big]
\end{eqnarray*}
where $|O (i) \rangle $ 
and  $|E (i) \rangle $ 
is a basis with odd (even) number of  spin up fermions respectively as 
\begin{eqnarray*}
|O(i) \rangle  =& 
 c_{m_1\uparrow}^\dagger  \cdots c_{m_{M_{u} }\uparrow} ^\dagger 
c_{m_1\downarrow}^\dagger  \cdots c_{m_{M_{d} }\downarrow} ^\dagger |0 \rangle 
\ ( M_u:{\rm odd,\ } M_d:{\rm even } ))
\\
|E(i) \rangle  =& 
 c_{m_1\uparrow}^\dagger  \cdots c_{m_{M_{u} }\uparrow} ^\dagger 
c_{m_1\downarrow}^\dagger  \cdots c_{m_{M_{d} }\downarrow} ^\dagger |0 \rangle 
\ ( M_u:{\rm even,\  } M_d:{\rm odd} ))
\end{eqnarray*}
They are 
orthonormalized as
\begin{eqnarray*}
\langle O(i) | O(j) \rangle = \langle E(i) | E(j) \rangle = \delta _{ij},
\quad
\langle O(i) | E(j) \rangle = 0
\end{eqnarray*}
where $i=1,\cdots,D_F$ is a label of the Fock states.
Since the total number of particles
is odd,
the basis with even up spins $|E(i) \rangle $ is given by that of the odd
as
\begin{eqnarray*}
|E(i) \rangle =& \Theta |O (i) \rangle .
\end{eqnarray*}
Therefore one has (also it is confirmed directly)
\begin{eqnarray*}
\Theta |E(i) \rangle =& \Theta^2| O(i) \rangle =-|O(i) \rangle 
\end{eqnarray*}
As for the generic state $|\psi \rangle $, the TR operation is given as
\begin{eqnarray*}
\Theta |\psi \rangle =& |\psi^\Theta \rangle =
\sum
(
-\psi_E^*(i)|O(i) \rangle 
+\psi_O^*(i)|E(i) \rangle 
)
\end{eqnarray*}

Using this set up, 
one can directly extend the discussion in the Sec.\ref{sec:trq}.
As for the eigen state
\begin{eqnarray*}
\psi =& \mvec
{\psi_O}{\psi_E},\quad
\psi_O=
\mvecthree{\psi_O(1)}{\vdots}{\psi_O(D_F)},
\psi_E=
\mvecthree
{\psi_E(1)}{\vdots}{\psi_E(D_F)},
\end{eqnarray*}
the KR multiplet of the many body state is given as
\begin{eqnarray*}
\Psi = (\psi,\Theta\psi)\equiv
\mmat
{\psi_O}{-\psi_E^*}
{\psi_E}{\psi_O^*}
=\psi^0 E+\psi^1 I+\psi^2 J+\psi^3 K
\end{eqnarray*}
where
 $\psi^0={\rm Re\,} \psi_O$,
 $\psi^1={\rm Im\,} \psi_O$,
 $-\psi^2=-{\rm Re\,} \psi_E$,
 $\psi^3={\rm Im\,} \psi_E$.
Orthogonality of the KR pair is also trivial.
Similar to the discussion in Sec.\ref{sec:trq}, we identity 
the KR multiplet as a single state of the quaternion as
\begin{eqnarray*}
\Psi  \cong\psi^{\mathbb{H}} =\psi^0 +\psi^1 \qi+\psi^2 \qj+\psi^3 \qk
\end{eqnarray*}
Then all of the discussion is trivially transformed into 
a discussion of the many body 
states.
For example, the quaternionic Berry connection for 
the many body state is defined as
$a^{\mathbb{H}} =(\psi^{\mathbb{H}} ) ^\dagger d\psi ^{\mathbb{H}} $.
All of the discussion in the paper can be applicable to the
many body system. Applications for electronic systems 
with electron-electron interaction will be given elsewhere. 

\section{ 
 Minimum dimensions for non-trivial
Berry connections}
\label{sec:minimum}
To have a non trivial topological structure in the Berry connection, 
there can be some requirements for the dimension of the parameter 
space $D$, which we describe here. 
Let us first start from a generic consideration of the
normalized $m$-dimensional  multiplet 
$\Psi=(\Psi_1,\cdots,\Psi_m)$, $\Psi ^\dagger \Psi=E_m$
and 
the corresponding $m$-dimensional non-Abelian 
Berry connection 
$
A = \Psi ^\dagger d \Psi= \Psi ^\dagger \partial_\mu  \Psi dx^\mu$,
 which transforms under a gauge transformation $\Psi_g=\Psi  g$,
$g\in U(m)$, as
$
A_g = g ^{-1} A g + g ^{-1} d g$\cite{Berry84,WZeeNA}.
The $n$-th Chern number $C_n$ of this connection is defined as
\begin{eqnarray*} 
C_n &=& \big(\frac {i}{2\pi} \big)^n \frac 1 {n!}
\int _{M_{2n}} \Tr F^n,
\ \
F = d A + A^2
\end{eqnarray*} 
where $M_{2n}$ is an $2n$-dimensional manifold without boundaries 
$\partial M_{2n}=0$\cite{Eguchi80,Zumino91}.
Although the field strength $F$ gets
 modified by the above gauge transformation
as $F_g= g ^{-1}  F g$, the Chern number is invariant.
As for the explicit discussion of the Berry connection,
 Zumino's generic construction of the topological quantities
is quite useful. We shortly summarize a part of them
which we require in this article\cite{Eguchi80,Zumino91}. 
They read
\begin{eqnarray*}
\Tr F &= d\omega _1 (A),\quad
\Tr F^2 &=  d\omega _3 (A),
\\
\omega _1(A) &= \Tr A, \qquad \omega _3(A) &=  \Tr( A dA+ \frac {2}{3} A^3)=\Tr ( AF-\frac {1}{3} A^3).
\end{eqnarray*}
The transformation properties are given as 
\begin{eqnarray*}
\fl\qquad  \omega _1(A_g) &=  \omega _1(A)  + \Tr ( g ^{-1} d g ),\quad 
\omega _3(A_g) = \omega _3(A)  -\frac {1}{3}  \Tr ( g ^{-1} d g )^3 
+ d \alpha _2
\end{eqnarray*}
where $ \alpha _2= \Tr (A dg g ^{-1})$. 
Although the Zumino's construction is general for 
$\Tr F^n = d\omega _{2n-1} (A)$, we just need for $n=1$ and $2$,
which one can explicitly confirm by a direct calculation. 

The Chern number is gauge invariant and it is 
explicitly given by the gauge invariant projection $P=\Psi\Psi ^\dagger $. 
It is given for the first Chern number\cite{AvronSum} but is also 
done for the higher ones.
By taking a differential of $P$, we have 
$dP=d\Psi \Psi ^\dagger  + \Psi  d\Psi ^\dagger $. Then 
the following direct calculation below
gives a useful formula for gauge invariant quantities
as
\begin{eqnarray*}
\Psi F \Psi ^\dagger = P dP^2 P,\quad 
{\rm Tr \,}( P dP^2 )^n = 
 {\rm Tr \,}F^n
\end{eqnarray*} 
It obeys from the following observation
\begin{eqnarray*}
 (dP)^2 &= 
d\Psi \Psi ^\dagger 
d\Psi \Psi ^\dagger 
+
\Psi d \Psi ^\dagger 
d\Psi \Psi ^\dagger 
+
d\Psi \Psi ^\dagger 
\Psi d \Psi ^\dagger 
+
\Psi d \Psi ^\dagger 
\Psi d \Psi ^\dagger 
\\
&=   
-d\Psi d\Psi ^\dagger 
\Psi \Psi ^\dagger 
+
\Psi d \Psi ^\dagger 
d\Psi \Psi ^\dagger 
+
d\Psi 
d \Psi ^\dagger 
-
\Psi  \Psi ^\dagger 
d\Psi d \Psi ^\dagger 
\\
&=  
-d\Psi d\Psi ^\dagger 
P
+
\Psi d \Psi ^\dagger 
d\Psi \Psi ^\dagger 
+
d\Psi 
d \Psi ^\dagger 
-
P
d\Psi d \Psi ^\dagger 
\\
P (dP)^2 P
&=  -Pd\Psi d\Psi ^\dagger 
P
+
P\Psi d \Psi ^\dagger 
d\Psi \Psi ^\dagger 
P
+
P d\Psi 
d \Psi ^\dagger 
P
-
P
d\Psi d \Psi ^\dagger 
P
\\
&=  
-Pd\Psi d\Psi ^\dagger 
P
+
P\Psi d \Psi ^\dagger 
d\Psi \Psi ^\dagger 
P
\\
&=  -\Psi \Psi ^\dagger d\Psi d\Psi ^\dagger 
\Psi \Psi ^\dagger 
+
\Psi \Psi ^\dagger \Psi d \Psi ^\dagger 
d\Psi \Psi ^\dagger 
\Psi \Psi ^\dagger 
\\
&=  
\Psi \Psi ^\dagger d\Psi \Psi ^\dagger 
d\Psi \Psi ^\dagger 
+
\Psi 
 d \Psi ^\dagger 
d\Psi
\Psi ^\dagger 
\\
&=  
\Psi \big[ d\Psi ^\dagger d \Psi + (\Psi ^\dagger d \Psi)^2\big] \Psi ^\dagger 
= 
\Psi F \Psi ^\dagger 
\end{eqnarray*}
where the normalization $\Psi ^\dagger \Psi = E_M$ implies 
$\Psi ^\dagger d\Psi=-d\Psi ^\dagger \Psi$, $P^2=P$
and
$dA =d\Psi ^\dagger d \Psi $.
Then the Chern number is written as an explicit gauge invariant form as
\begin{eqnarray*}
\fl\qquad \quad
C_n = \big(\frac {i}{2\pi} \big)^n\frac {1}{n!} \int_{M_{2n}} \Tr\big[ P (dP)^2P\big]^n=
 \big(\frac {i}{2\pi} \big)^n\frac {1}{n!} \int_{M_{2n}} \Tr\big[ P (dP)^2\big]^n.
\end{eqnarray*}

As for the TR invariant system with the KR degeneracy,
 we  identify the multiplet of the 
dimension $2M$ to the quaternionic one with the dimension $M$ as 
$\Psi=(\Psi_1,\cdots,\Psi_M)\cong\psi ^{\mathbb{H}} $.
Then 
a gauge transformation $\psi_g ^{\mathbb{H}}  =\psi ^{\mathbb{H}} g$, 
$g\in Sp(M)$  preserves  the TR invariant  linear  space 
spanned by $\psi ^{\mathbb{H}} $. 
Now the quaternionic
 Berry connection $a ^{\mathbb{H}} 
  =(\psi^{\mathbb{H}} )  ^\dagger d \psi^{\mathbb{H}}    $  and corresponding 
field strength $f^{\mathbb{H}}  = d a^{\mathbb{H}}  + (a^{\mathbb{H}} )^2 $ are 
defined as usual. Their transformation properties are also standard as
$a^{\mathbb{H}} _g =
(\psi^{\mathbb{H}}_g )^\dagger d \psi^{\mathbb{H}} _g
=g ^{-1} a ^{\mathbb{H}} g + g ^{-1} d g$ and
$ f _g^{\mathbb{H}} 
=d a^{\mathbb{H}} _g + (a_g^{\mathbb{H}} ) ^2 = g ^{-1} f^{\mathbb{H}}  g$.
The $n$-th Chern  number with {\em even} $n$, $C_n$ is defined as 
(
Since the $C_n$ is 
intrinsically integer, it suggests vanishing $C_n$ for odd $n$)
\begin{eqnarray*} 
\fl \qquad  C_n &=  \bigg(\frac {-1}{4\pi^2} \bigg)^{n/2} \frac 1 {n!}
\int _{M_n}  \Tr _M T\, (f^{\mathbb{H}} )^n  
=
 \bigg(\frac {-1}{4\pi^2} \bigg)^{n/2} \frac 1 {n!}
\int _{M_n}  \Tr _M T \, [p^{\mathbb{H}} (dp^{\mathbb{H}} )^2 ]^{n/2}
\end{eqnarray*} 
where 
$T x ^{\mathbb{H}} =x+\bar x=2 x^0\in \mathbb{R}$ for a quaternion $x = 
x^0+ x^1\qi+x^2\qj+x^3\qk $ and 
the quateronic projection is 
 $p^{\mathbb{H}} =\psi^{\mathbb{H}} (\psi^{\mathbb{H}}) ^\dagger $.
In the following, we omit the symbol $^{\mathbb{H}} $ and simply
use the lower character for the quaternionic notation if the situation is 
clear. 

Since the multiplet and the Berry connection have a gauge freedom, one needs 
to fix it
for the connection to be well-defined.
As for the generic multiplet without the KR degeneracy, 
the gauge is specified by an arbitrary but given multiplet $\Phi$
as 
$\Psi_\Phi=P\Phi N_\Phi^{-1/2}$ where
$P$ is a gauge independent projection and the normalization
matrix
 $N_\Phi=\Phi ^\dagger P \Phi$ which is also gauge invariant and
is semi-positive definite\cite{Hatsugai04e}.
When one can use this single gauge over the whole parameter space, the 
Berry connection is trivial.  
Generically, however, the normalization matrix may have zero
eigen values as  $\det N_\Phi(^\exists x_\Phi)=0$. Then near this zero, $x_\Phi$,
this gauge is singular since one can not normalize.
One needs to use the other gauge, say, 
$\psi_{\Phi^\prime}$ by taking $\Phi^\prime$. 
Since
 $\det N_{\Phi^\prime}(x_\Phi)\ne 0$, generically,
one can express the projection by the multiplet explicitly as
$P=\Psi_{\psi^\prime}\Psi_{\psi^\prime} ^\dagger $ and 
 the normalization matrix is factorized 
as
$N_{\Phi}=\Phi ^\dagger P\Phi= \eta_{\Phi^\prime\Phi} ^\dagger \eta_{\Phi^\prime\Phi}$,
$\eta_{\Phi^\prime\Phi}\equiv \Psi_{\Phi^\prime} ^\dagger \Phi $.
One may write it as $\eta_{\Phi}=\Psi ^\dagger \Phi$ 
when one does not need to
specify the gauge.
Now it is clear that the  singularity is specified by
\begin{eqnarray*}
\det \eta_{\Phi^\prime\Phi}
 =0
\rightleftarrows
 {\rm Re\,}\big( \det \eta_{\Phi^\prime\Phi}\big)
=
 {\rm Im\,}\big( \det \eta_{\Phi^\prime\Phi}\big)=0
\end{eqnarray*}
since 
this determinant is complex, 
$\det\eta_\Phi(x)\in \mathbb{C}$.
Generically one does not 
have zeros when the dimension of the 
parameter space is too low and the Berry connection is
trivial.
To have a non trivial topological structure
 the dimension of the parameter space has to satisfies $D\ge D_{min}=2$,
since the condition to have the singularities is given by
the two real equations.
A two-dimensional magnetic Brilluine zone to discuss the Hall conductance 
as the first Chern number is this minimiun space where the singularities
 occur in points\cite{Kohmoto85}.
Note that the gauge transformation between the two gauges 
by 
$\Phi$ and $\Phi^\prime$,
$\Psi_{\Phi^\prime}=\Psi_\Phi g_{\Phi\Phi^\prime}$,
 is explicitly given by
\begin{eqnarray*}
\Psi_{\Phi} &= 
\Psi_{\Phi^\prime}\Psi_{\Phi^\prime} ^\dagger \Phi N_{\Phi}^{-1/2}=
\Psi_{\Phi^\prime}g_{\Phi^\prime\Phi},
\\ 
g_{\Phi^\prime\Phi} &= 
\Psi_{\Phi^\prime} ^\dagger \Phi N_{\Phi}^{-1/2}=
(N_{\Phi^\prime})^{-1/2}
{\Phi^\prime} ^\dagger P \Phi (N_{\Phi})^{-1/2}
\in U(M).
\end{eqnarray*}
The  unitarity is confirmed as 
\begin{eqnarray*}
\fl\quad 
g_{\Phi^\prime\Phi}g_{\Phi^\prime\Phi} ^\dagger = 
(N_{\Phi^\prime})^{-1/2}
{\Phi^\prime} ^\dagger P \Phi 
 (N_{\Phi})^{-1}
{\Phi} ^\dagger P \Phi^\prime
(N_{\Phi^\prime})^{-1/2}
\\
\fl \qquad = 
(N_{\Phi^\prime})^{-1/2}
\eta_{\Phi^\prime} ^\dagger \eta_\Phi
(N_\Phi) ^{-1} \eta_{\Phi} ^\dagger \eta_{\Phi^\prime}
(N_{\Phi^\prime})^{-1/2}
= 
(N_{\Phi^\prime})^{-1/2}
\eta_{\Phi^\prime} ^\dagger
 \eta_{\Phi^\prime}
(N_{\Phi^\prime})^{-1/2}
=E_M
\end{eqnarray*}
and  $g_{\Phi^\prime\Phi} ^\dagger g_{\Phi^\prime\Phi}=E_M$ similarly.

As for a systems with the KR pairs,
 let us here consider the simpest $M=1$ case.
Now starting from the gauge invariant projection $p$ into the
degenerate KR space, the gauge is fixed by
an aribitrary quaternion vector $\phi\in \mathbb{H}^N $ as
\begin{eqnarray*}
\psi_\phi
=
 p
 \phi N_\phi^{-1/2},\quad
 N_\phi=\phi  ^\dagger p \phi
=N(\eta_{\phi})\in\mathbb{R},
\quad \eta_{\phi}
=\psi^\dagger \phi \in \mathbb{H}
\end{eqnarray*}
where
 $N(x)=\bar x x=
(x^0)^2
+
(x^1)^2
+
(x^2)^2
+
(x^3)^2
$ is  a norm of a quaternion $x\in \mathbb{H}$.
This gauge is again well defined only if $ N_\phi\ne 0$.
Note that although $\eta_\phi$ itself is gauge dependent but 
the norm $N(\eta_{\phi})$ is gauge invariant as
 $N(\psi_g ^\dagger \phi)=N(\bar g\psi ^\dagger \phi)=N(g)N(\psi ^\dagger \phi)
=N(\psi ^\dagger \phi)$, 
($\psi_g=\psi g$, $g\in Sp(1)$).
Therefore we do not need to specify the gauge for $N(\eta_\phi)$.

Near the singular point of this gauge, one needs to use the other gauge by
${\phi^\prime}$. Then the condition of the vanishing norm
$N_\phi=N(\eta_{\phi^\prime\phi})$,
 is expressed as
\begin{eqnarray*}
\eta_{\phi^\prime\phi}
= 0 \rightleftarrows
T\big(\eta_{\phi^\prime\phi} \big)
=
T\big(\qi \eta_{\phi^\prime\phi} \big)
=
T\big(\qj\eta_{\phi^\prime\phi} \big)
=
T\big(\qk\eta_{\phi^\prime\phi} \big)=0.
\end{eqnarray*}
It clearly shows that the singularity may occur in the parameter space 
of the dimension $D\ge D_{min}^{\rm KR}=4$.
The gauge transformation is also give as
\begin{eqnarray*}
\psi_\phi 
= 
\psi_{\phi^\prime}  g_{\phi^\prime\phi},
\ \
g_{\phi^\prime\phi}
=
[ N({\phi^\prime})]^{-1/2}
({\phi^\prime}) ^\dagger p
\phi
[ N(\phi)]^{-1/2}\in Sp(1).
\end{eqnarray*}

When the dimension of the parameter space is less than this minimum dimension, 
one can generically take a single patch over the whole
parameter space. 
Since the base space to define the Chern numbers are assumed to 
be without boundaries, 
it implies that the Chern number is vanishing for 
${\rm dim}\, M_{2n}=2n<D_{min}^{\rm KR}=4$. 
Then the natural quantities to have non trivial
 topological structure by the Chern numbers are
$C_1$ for the generic case and $C_2$ for the system with the KR degeneracy.

Also note 
 that
the normalization  
of the KR pair in quaternion notation
 $\psi ^\dagger \psi=1$ gives
$0
={\psi} ^\dagger d {\psi}+d{\psi} ^\dagger {\psi} 
={\psi} ^\dagger d {\psi}+\widetilde {d{\psi} ^\dagger {\psi} }
={\psi} ^\dagger d {\psi}+\widetilde {{\psi}} d\bar {\psi} =
T({\psi} ^\dagger d {\psi})=T (a)
$, which implies the first Chern number is vanishing
that is consistent with the generic argument 
\cite{Avron88,Avron89}.
Then let us focus on the 2nd Chern number with the KR degeneracy. 

\section{Degeneracies to  Monopoles with/without KR degeneracy}
\label{sec:deg}
As was pointed out by Berry,
the generic degeneracy of a
complex hamiltonian has a co-dimension $d_{C}=3$\cite{Berry84}, 
that is, 
the minimum hamiltonian ($N=2$)
to describe the degeneracy (at $E=\Tr H_{\mathbb{C}}=0$) is a complex 
hermite  $2\times 2$ matrix $H_{\mathbb{C}}$ which is 
expanded by the Pauli matrix with 3 dimensional real coefficients
$\bi{R} (x)=( R_1(x),R_2(x),R_3(x))\in\mathbb{R}^3$
as 
\begin{eqnarray*}
H_{\mathbb{C}}(x)=
\mmat
{R_3}{z}
{\bar z}{-R_3},\quad z=R_1-iR_2
\end{eqnarray*}
where $R_3=R_3(x)\in \mathbb{R}$, $z=z(x)\in \mathbb{C}$.
Similarly 
the  system with the KR degeneracy does have a co-dimension  $d_{H}=5$
as pointed out by Avron et al.\cite{Avron88,Avron89}. 
Then the minimum model ($N=2$,  $E=\Tr H_{\mathbb{H}}=0$)
is realized by the following quaternionic hermite hamiltonian 
\begin{eqnarray*}
H_{\mathbb{H}}(x)
&=&  \mmat
{Q_5}{q}
{\bar q}{-Q_5},\quad q=q_0+q_1\qi +q_2 \qj
 +q_3 \qk
\end{eqnarray*}
where $Q_5=Q_5(x)\in \mathbb{R}$,
$q_i(x)\in \mathbb{R}$, $(i=1,2,3)$ and
 $q=q(x)\in\mathbb{H}$.
These $\bi{Q}=(Q_1,Q_2,Q_3,Q_4,Q_5)\in \mathbb{R}^5$,
($Q_1=q_1,Q_2=q_2,Q_3=q_3,Q_4=q_0$) form
 5 dimensional parameters of the minimum model with the KR degeneracy. 

\begin{figure} [h] 
\begin{center} 
\includegraphics[width=8cm]{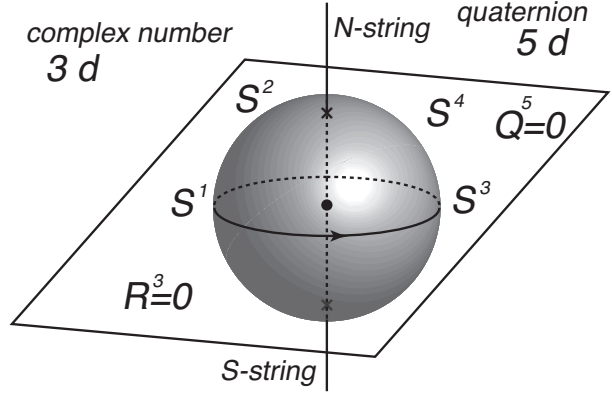}             
\caption{Topological objects and singularities 
for the Dirac monopole and the Yang monopole.}
\label{fig:sphere}
\end{center} 
\end{figure}

The above observation suggests strong analogy between the systems with and 
without the KR degeneracy,
 which we pursuit in this paper. 
There is also a topological correspondence as discussed below
(See Fig.\ref{fig:sphere}).
 Actually it is 
more than analogy and there exists a mapping by
$R_3 \to
Q_5
$
and
$z(\in\mathbb{C})\to q(\in \mathbb{H})$
as one can see.
The origins of the parameter spaces $\bi{R}=0 $ and $\bi{Q}=0 $ give
 degeneracies which bring singularities for the each Berry connections.
They are the Dirac monopole\cite{dirac} and the Yang monopole\cite{Yang78,Minami79,Minami80,Demler99,Murakami04,BHZ,FuKane06}.
The Yang monopole is literally a quaternionic Dirac monopole up to 
its topological structure.
 
\subsection{Dirac monopole and the first Chern number}
\label{sec:dirac}
Due to a simple observation, 
$ H_{\mathbb{C}} ^2=R^2 E_2$,  $R=|\bi{R}|$, 
the energies of $ \hc  $ are $\pm R=\sqrt{|z|^2+R_3^2}$.
Then the degeneracy 
occurs at the origin in the 3 dimensional $\bi{R}$ space $\rthree$.
Away from this degeneracy, the eigen state of the
energy $\pm R$ subspace is well
defined by the projection $P_\pm   =
\case 1 2  (1\pm H_{\mathbb{C}} /R)$.
As for the base manifold to define the first Chern number, 
for simplicity, let us take 
the 2-sphere $S^2=\{\bi{R}| R=1\}\subset \rthree$ as for 
$M_{2n}$, $n=1$ (Fig.\ref{fig:sphere}).
Then the possible singularities of the Berry connection
can be  points on the $S^2$ by the generic consideration before.
When one considers a generic base space in $\rthree$, these singularities
form  lines, which correspond to the Dirac strings\cite{wu-yang}.
The gauge invariant projection into each eigen subspace is explicitly
given as $P_\pm=\case 1 2 \mmat{1\pm R_3}{\pm z}{\pm \bar z}{1\mp R_3}$.
In the following, let us consider a positive energy subspace $P=P_+$.
Taking a gauge
by $\Phi_N=\mvec{1}{0}$,
the normalized state on $S^2$, ($|z|^2+R_3^2=1$), is given as
 a $\Psi_{N}=P \Phi_{N} N_N^{-1/2}$ with 
$N_{N}= \Phi_N ^\dagger  P \Phi_{N}= \case 1 2 (1+R_3)$.
Since this gauge is only  singular at the south pole $R_3=-1$, 
we can safely use 
$
\Psi_{N}=
\case 1 {\sqrt{2}}
\mvec
{(1+R_3)^{+1/2}}
{\bar z (1+R_3)^{-1/2}}
$
 for the north hemisphere $S^2_N$ ( $R_3\ge 0$). 
As for the south hemisphere, we needs to use the other gauge, say, 
by $\Phi_S=\mvec{0}{1}$. Then the normalized state is given similarly as
$\Psi_{S}=P_{\mathbb{C} }\Phi_{S} N_S^{-1/2}=
\case 1{\sqrt{2}}
\mvec
{z(1-R_3)^{-1/2}}
{(1-R_3)^{+1/2}}
$, 
$N_{S}= \Phi_S ^\dagger  P_{\mathbb{C} }\Phi_{S}= \case 1 2 (1-R_3)$,
which is regular everywhere on the south hemisphere $S^2_S$ ( $R_3\le 0$).

The gauge transformation, $g^{\mathbb{C}}_{SN} $,
 between them,
$ \Psi_N=\Psi_S g^{\mathbb{C}}_{SN}$,
 is given by the generic formula before
as 
\begin{eqnarray*}
g^{\mathbb{C}} _{SN}
&=& N_{S}^{-1/2}\Phi_S ^\dagger  P_{\mathbb{C}}\Phi_N N_N^{-1/2}=\bar z/|z|.
\end{eqnarray*}
This is regular except the north and south poles $R_3=\pm 1$.

The first Chern number 
of the Berry connection is easily evaluated 
using these two gauges and the gauge transformation,
$
A_N=g_{SN}^{-1} A_S g_{SN}+g_{SN}^{-1} d g_{SN}
$
\begin{eqnarray*}
\fl 
\qquad C_1 &=  \frac {i}{2\pi} \int_{S^2} \Tr F=
 \frac {i}{2\pi} \int_{S^2} d \omega _1(A)
= 
\frac {i}{2\pi} \bigg(
\int_{S^2_N} d \omega _1( A_N) +
\int_{S^2_S} d \omega _1(A_S)
\bigg)
\\
\fl &= 
\frac {i}{2\pi} \bigg(
\int_{ \partial S^2_N} \omega _1(A_N) +
\int_{\partial S^2_S}  \omega_1( A_S)
\bigg)
= 
\frac {i}{2\pi}
\int_{ S^1=\partial S^2_N} \big(\omega _1(A_N)-\omega _1(A_S)\big)
\\
\fl &=  W_{S^1}(g_{SN}^{\mathbb{C}})
\end{eqnarray*}
where 
$S^1=\partial S^2_N=\partial S^2_S$ is an equator $S^1=\{\bi{R}|R=1, R_3=0\}$
and 
 $W_{S^1}(g_{SN}^{\mathbb{C}} )$ 
 is a winding number of the map from the 1-sphere (circle)
 $S^1=\{(R_1,R_2)|R_1^2+R_2^2=1\}$
to $U(1)\cong S^1=\{z\big||z|^2=1\}\in \mathbb{C}$ as
\begin{eqnarray*}
 W_{S^1}(g_{SN}^{\mathbb{C}}) =
\frac {i}{2\pi}\int_{S^1}
(g^{\mathbb{C}}_{SN}) ^{-1} 
d g^{\mathbb{C}}_{SN}
=-1.
\end{eqnarray*}
This winding number
 can be evaluated by several ways. Since this is invariant against a
rotation in $S^1$ ( $g\to e^{i\theta} g$ ), we write it in a local coordinate 
near $R_1=0$ and $R_2=1$ as $g=-i $, $d g = d R_1 $ as 
$ W_{S^1}(g_{SN}^{\mathbb{C}}) =(i/2\pi)  \int _{S^1}d R_1/(-i)
=-\int _{S^1}d R_1/(2\pi)=-1$
 where $\int _{S^1}d R_1=2\pi$ is a volume ( length )
of the circle $S^1$.
Also using the explicit form $g_{SN}^{\mathbb{C}}=e^{i{\rm Arg}\, (R_1+i R_2)} $,
we have $\int _{S^1} g ^{-1} d g =
i\int _{S^1}  d\,  {\rm Arg} \,  (R_1+i R_2) =2\pi i$. 

Considering the $S^2$ as a boundary of the solid sphere $V_3$
($ \partial V_3=S^2$), 
naive application of the Stokes (Gauss) theorem, 
$ \displaystyle C_1=\int_{V_3} d F$, suggests
$\frac {i}{2\pi} dF = -\delta^{(3)}( \bm{R}) $
since $dF=d^2 A=0$ as far as the Berry connection is well-defined
except the origin.
This is the {\em Dirac monopole } at the origin of the 
3-dimensional $\bm{R} $  space 
where the degeneracy of the generic complex hamiltonian occurs\cite{dirac}.

\subsection{Yang monopole as a quateronic Dirac monopole}
\label{sec:yang}
The discussion with the KR degeneracy can be done quite analogously.
Let us again start from a simple observation $H_{\mathbb{H} }^2 =Q^2 E_5$, 
$Q=|\bi{Q}|$, which implies that eigen energies  of the KR multiplets
are $\pm Q=\pm\sqrt{|q|^2+Q_5^2}$, $|q|=\sqrt{N(q)}\in \mathbb{R}$,
 and the 
additional degeneracy to the Kramers degeneracy occurs at the origin
 in the 5-dimensional $\bi{Q}$ space $\rfive$ (Fig.\ref{fig:sphere}). 
A projection into the 
positive energy KR  multiplet is defined as
$ p =\frac 1 2 (1+ H_{\mathbb{H}}/Q)$.
 Similar to the discussion above, 
let us take a 4-sphere $S^4=\{\bi{Q}\big|Q=1\}\subset \rfive$
as the base space $M_{2n}$,
 $(n=2)$ to define the second Chern number $C_2$. 
Then the generic singularities of the KR multiplet are again points 
on $S^4$, which make lines in the $\rfive $ when one considers a generic 
4 dimensional surface as a base space ("Yang'' strings).
To be more specific, let us take a gauge by 
taking a quaternion vector with 2 components
 $\phi_N=
\mvec{1}{0}\in \mathbb{H}^2$. 
Then the normalized KR multiplet is given, in the north pole gauge
(regular in the north hemisphere $S^4_N(Q_5\ge 0)$ ), as
\begin{eqnarray*}
\psi_N &=& p \phi_N N_N^{-1/2}
=\frac 1 {\sqrt{2}}\mvec
{(1+Q_5)^{+1/2}}
{\bar q (1+Q_5)^{-1/2}}
\end{eqnarray*}
where $N_N= \phi_N ^\dagger p \phi_N
= \case 1 2 (1+Q_5)$.  This gauge is only singular at the south pole $Q_5=-1$
on the $S^4$. 
The other gauge by
 $\phi_S =
\mvec{0}{1}$ also defines the multiplet ( in the south pole gauge )
\begin{eqnarray*}
\psi_S &=& p \phi_SN_S^{-1/2}
=\case 1 {\sqrt{2}}\mvec
{q(1-Q_5)^{-1/2}}
{(1-Q_5)^{+1/2}}
\end{eqnarray*}
where $N_S= \phi_S ^\dagger p \phi_S= \case 1 2 (1-Q_5)$.
 This is regular in the south hemisphere $S^4_S(Q_5\le 0)$.
The gauge transformation between them is also calculated as
\begin{eqnarray*}
 \psi^{\mathbb{H}} _S&=&  g^{\mathbb{H}} _{SN}  \psi^{\mathbb{H}} _N,\quad 
g^{\mathbb{H}} _{SN} = \bar q /|q|\in Sp(1)=\{g\in \mathbb{H}|N(g)=1\}
\end{eqnarray*}
Now let us calculate the second Chern number in the quaternionic notation
as
\begin{eqnarray*}
\fl\qquad   C_2 &=  -\frac {1}{8\pi^2} \int_{S^4} T f^2
= -\frac {1}{8\pi^2} \int_{S^4}  d \omega _3(a)
= -\frac {1}{8\pi^2}\bigg(
 \int_{S_N^4}  d \omega _3(a_N)
+
 \int_{S_S^4}  d \omega _3(a_S)\bigg)
\\
\fl  &=  -\frac {1}{8\pi^2}
 \int_{S^3}  \big( \omega _3(a_N) -\omega _3(a_S)\big)
=  \frac {1}{24\pi^2} \int_{S^3} 
 T( (g_{SN}^{\mathbb{H}} ) ^{-1}  d g_{SN}^{\mathbb{H}}   )^3
\\
\fl & \equiv 
W_{S^3} (g_{SN}^{\mathbb{H}} )=-1
\end{eqnarray*}
where  $S^3=S^4\big|_{Q_5=0}=\{(q_1,q_2,q_3,q_0)\big| |q|=1\}$ is an equator,
$\omega _3( a ) = T( a da + \frac {2}{3} a^3 )$ and  
$W_{S^3} (g_{SN}^{\mathbb{H}} )$ is the 
Pontrjagin number of the map 
$S^3\to Sp(1)\cong S^3$ that is a covering degree, which is 
intrinsically integer.
Here we used $\int_{S^3} d \alpha _2=\int_{\partial S^3}  \alpha _2=0$
since the gauge is regular on the $S^3$ which does not have boundaries.
This Pontrjagin number is explicitly evaluated\cite{Bertl}.
Since it is invariant for the change $q\to q \xi $, $|\xi|=1$
that induces a rotation of $S^3$, 
it is enough to evaluate it near $q=1$
  ($q_0=1, q_1=q_2=q_3=0$), 
where $T(q ^{-1} dq)^3=3!T(\qi\qj\qk) dq_1dq_2dq_3=-12 dq_1dq_2dq_3$.
Then we have $C_2(\bm{Q})= W_{S^3}(g_{SN}^{\mathbb{H}} )
=\frac {1}{24\pi^2} (-12\cdot 2\pi^2)=-1 $ where $2\pi^2$ is a volume
of the $S^3$. 

Again writing the $S^4$  as a surface of 
a 5 dimensional solid sphere $V_5=\{\bm{Q}\big||\bm{Q} |\le 1\} $, 
$\partial V_5=S^4$, one may write symbolically
$
d T( f^2) = - \delta ^{(5)}( \bm{Q} )
$ by a simple application of the Stokes theorem
$\int _{V_5} d T(f^2) =
\int_{\partial V_5} T (f^2) =-1
$,
since $d\,  T(f^2)=d^2 \omega _3=0$ away from the origin where the 
singularity exists.
The origin of the 5-dimensional $\bm{Q} $ space,
 $\bm{Q} =0$, is a singular point for the Berry connection due to the 
additional degeneracy (4-fold)
and it induces  the Yang's monopole in 5-dimensions\cite{Yang78} which
locates at $\bm{Q}=0 $
(the charge is $-1$).
This explicit demonstrates that
{\em the Yang monopole is a quaternioninc Dirac monopole.}

%

\subsection{Chiral symmetry and topological stability of the Dirac cones}
\label{sec:chiralTop}
For simplicity, 
we have assumed the 2-sphere and the 4-sphere
as the parameter spaces $M_2 $ and $M_4$.
In a generic situation, let us consider the Chern numbers of the  models
$\hc(\bi{R}(x))$, ($x\in M_2$)  and $\hq(\bi{Q}(x))$ ($x\in M_4$).
Assuming the energy 
gaps never collapses, 
 the images $\bi{R}\bi(M_2)\subset \mathbb{R}^3$  and $\bi{Q}(M_4)\subset\mathbb{R}^5$ are 
deformed into the spheres $S^2$ and $S^4$ without changing the
Chern numbers.  This is the topological stability and 
these topological numbers are given 
by the covering degrees of the maps as\cite{Hatsugai02sup}
\begin{eqnarray*}
\fl\qquad \quad 
C_1 =- {\rm deg}\, \bi{R}(M_2)\, :M_2\to S^2,\quad 
C_2 =- {\rm deg}\, \bi{Q}(M_4)\, :M_4\to S^4.
\end{eqnarray*}

\begin{figure} 
\begin{center} 
\includegraphics[width=10cm]{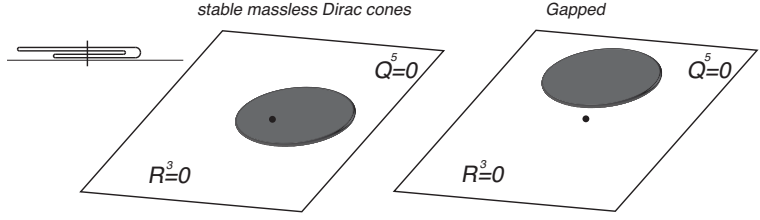}             
\caption{Collapsed images of the maps into the hyperplanes
 $M_2\to \bi{R}\subset \mathbb{R}^2:(R_3=0)$
and
$M_4\to \bi{Q}\subset \mathbb{R}^4:(R_3=0)$
with the chiral symmetric minimum models.
}
\label{fig:collapsed}
\end{center} 
\end{figure} 

To have the well-defined Chern numbers, the gap has to be open always. 
However in some situation, the gap may collapse. 
Generically speaking, this is accidental (accidental degeneracy).
In other words, one may need to fine tune physical parameters which
occurs at 
a quantum critical point.
By imposing  some restriction by symmetry, the situation may change and the
gap closing has a topological stability.
Let us here impose a "chiral symmetry'' and 
restrict the parameter space.
The chiral operator in the minimum model is given by  $\Gamma =\sigma _3$,
$\Gamma ^2=1$.
The hamiltonians of the minimum models satisfy
$\{\hc,\Gamma \}= 2R_3$, $ \{\hq, \Gamma \} =2 Q_5$. That is, the equators
($S^1$  and $S^3$ respectively)
are characterized as the chiral symmetrical spaces
\begin{eqnarray*}
\{\hc(\bi{R}),\Gamma \}= 0\ (\bi{R}\in S^1), 
\quad \{\hq(\bi{Q}), \Gamma \} =0\ (\bi{Q}\in S^3)
\end{eqnarray*}
When the hamiltonians do have the chiral symmetry, the parameter spaces
$\bi{R}(M_2$ (for $\hc$) and $\bi{Q}(M_4)$ (for $\hq$) 
are collapsed into the hyperplane  $\mathbb{R}^2(R_3=0)$ and 
 $\mathbb{R}^3(Q_5=0)$.
 Then we have two situations for the images $\bi{R}(M_2)$/$\bi{Q}(M_4)$
(See Fig.\ref{fig:collapsed}). 
The one case is that the   $\bi{R}(M_2)$/$\bi{Q}(M_4)$
includes the origin and in the other case, it does not.
 When the image includes the
origin, it implies the energy gap collapses and is the gap 
is linearly vanishing as a function of the parameter $x$ generically.
It brings  a Dirac-cone like 
energy dispersion. 
The doubling is also topologically clear (See the inset of the
Fig.\ref{fig:collapsed}). 
This Dirac cones are
generically topologically stable, that is, stable against for small but 
finite perturbation
since the images   $\bi{R}(M_2)\subset \mathbb{R}^2$
and $\bi{Q}(M_4)\subset \mathbb{R}^4$.
These topological stability of the Dirac cones in $2$/$4$ dimensions
are discussed in relation to the graphene and Nielsen-Ninomiya theorem
\cite{HFAstability,GraCreutz,NNT}.

\section{
Symmetry protected $\mathbb{Z}_2$ quantization 
}
\label{sec:z2}
As discussed,
the Chern numbers are gauge invariant and intrinsically integer 
which apparently have a topological stability. It implies that the
quantization
is stable for small but finite perturbation for the hamiltonian. 
This topological stability does play a crucial role, for example, 
in the theory of the quantized Hall effects.
Note that the dimension of the parameter space to
define the Chern numbers is necessarily even. 
The winding number $W_{S^1}$ and the 
Pontrjagin index $W_{S^3}$ are also topological as their definitions
and defined for the spaces with odd dimensions. 
Further in odd dimensions, 
one may also define quantized quantities
if one imposes additional symmetry requirements.
They are generalizations of the Berry phase and are 
generically gauge dependent
as a phase of the wave function\cite{Berry84,Hatsugai06a}.
It implies these quantities are essentially quantum mechanical
and do not have any classical correspondents.
They also
have a fundamental advantage 
in the identification of the topological ordered 
states\cite{Hatsugai05-char,Hatsugai06a}.
An example is a $\mathbb{Z}_2$-quantization of the  Berry phase for 
the TR invariant system without the KR 
degeneracy $\Theta^2=1$ \cite{Hatsugai06a,HiranoDeg,HiranoTopo,MaruSpinBerry}. 
The focus of this section is to extend the idea and 
supplies a generic condition for the $\mathbb{Z}_2$-quantization.

Now let us start by defining  generic 
Berry phases $\gamma_1(A) $  and 
$\gamma _3(a )$
as
\begin{eqnarray*} 
\gamma _1(A)  = \frac{i}{2\pi} \int_{S^1} \omega _1(A), \quad
\gamma _3(a )  =  -\frac{1}{8\pi^2}
 \int_{S^3} \omega _3(a)
\end{eqnarray*} 
where $\gamma _1(A)$ is for a generic system (without degeneracy $M=1$)
 and 
$\gamma _3(a )$ is for a system with the KR degeneracy
using a quaternionic notation before. 
Note here that the 
 same topological quantity by the integral of the 
Chern-Simons form is discussed in several papers\cite{NiuCS,Qi08,EssinMooreVan}.
They are not invariant
for the gauge transformation
 $A_g= g ^{-1} A g + g ^{-1} d g $ ($g\in U(1)$)
and   $a_g= g ^{-1} a g + g ^{-1} d g $  ($g\in Sp(1)$).
Therefore they are not well defined (as they are)
but are gauge independent
and well-defined
in modulo 1 as\cite{Hatsugai06a} 
\begin{eqnarray*}
\fl\qquad
\gamma _1(A_g)  =   \gamma _1(A)   +W_{S^1}(g)
\equiv \gamma _1(A), 
\quad
\gamma _3(a_g)  =   \gamma _3(a)   +W_{S^3}(g)
\equiv \gamma _3(a)
\end{eqnarray*}
since the gauge dependence is due to a non trivial large gauge
transformation. These contribution are
topological and integers
as $W_{S^1}(g)\in \mathbb{Z}$
 and $W_{S^1}(g)\in \mathbb{Z}$\cite{Hatsugai06a}
as far as the gauge transformations are regular over the $S^1$ and $S^3$.
A phase factor of the 
Berry phase $e^{i2 \gamma }$ ($\gamma =2 \pi \gamma _1$) 
is gauge independent 
and is a well defined quantity (observed as a geometrical  phase)
 but the phase $\gamma $ itself
is gauge dependent\cite{Berry84,Hatsugai06a}. 

Generically speaking, 
these generic Berry phases $\gamma _1$ and $\gamma _3$ may take any 
real values even in modulo 1.
 However they can be quantized when the system obey some symmetry
requirement which we discuss below.

\subsection{$\mathbb{Z}_2$ quantization of TR invariant system without KR degeneracy}
\label{sec:tr}

Let us first consider a TR invariant system 
without the KR degeneracy\cite{Hatsugai06a,HiranoTopo,HiranoDeg,MaruSpinBerry}.
This is realized for  quantum  systems with even number of quantum spins. 
Since the hamiltonian ${\cal H}$ does commute with the TR operator $\Theta$, 
which is anti-unitary, 
$[{\cal H},\Theta] =0$
\begin{eqnarray*}
\qquad \fl  {\cal H}(x)\psi(x) = \epsilon (x)\psi(x),
\quad {\cal H}(x)\psi^\Theta(x)= \epsilon(x) \psi^\Theta(x),
 \quad \psi^\Theta\equiv \Theta\psi
\end{eqnarray*}
Due to the uniqueness of the state, $\psi$ and $\psi^\Theta$ are 
only different in phase, that is, the corresponding Berry connections
$A=\psi ^\dagger d\psi$ and $A^\Theta=(\psi^\Theta) ^\dagger d\psi^\Theta$ 
are transformed each other by some gauge transformation $g$, 
$A=g ^{-1} A^\theta g +g ^{-1} d g$,
as $
\gamma _1(A) \equiv \gamma _1(A^\Theta)\ {\rm mod}\, 1
$, 
since the gauge transformation is, generically,  well defined on
the parameter space $x\in S^1$. 
Also the time reversal operation for the
many spin state $\psi$ is written as $\Theta =U {\cal K}  $ with some 
parameter independent unitary transformation $U$.
Then the Berry connection is written as
\begin{eqnarray*}
A^\Theta = (\psi^\Theta) ^\dagger d \psi^\Theta =
 {\cal K} A =-A
\end{eqnarray*}
since the normalization $\psi ^\dagger \psi=1$ implies 
that
$0=(d\psi ^\dagger) \psi+\psi ^\dagger d \psi
=\widetilde{d\psi ^\dagger \psi}+A
=\widetilde{\psi}d\psi ^*+A=A^*+A$.
Now we have two conditions for the Berry phases
\begin{eqnarray*}
\gamma _1(A) \equiv \gamma _1(A^\Theta)=-\gamma _1(A)\quad {\rm mod}\, 1
\end{eqnarray*}
Therefore  allowed values of the Berry phase are restricted into 
two as $\gamma _1(A)=0,\case 1 2 $.
This is the $\mathbb{Z}_2$-quantization of the Berry phase for
the unique TR invariant state. 

In most of the 
application\cite{Hatsugai06a,HiranoTopo,HiranoDeg,MaruSpinBerry}, 
we have used a $U(1)$ twist $e^{i\theta}$,
$\theta:0\to 2\pi$ as a parameter. In this case, the condition of 
the $\mathbb{Z}_2$-quantization is reformulated from a more 
generic point of view (See below). 

\subsection{$\mathbb{Z}_2$-quantization by inversion/reflection equivalence}
\label{sec:inv}

Similar quantizations protected by symmetry 
occur for 
the generic Berry phases, $\gamma _1$ and $\gamma _3$, 
when  the system (with parameter) does satisfy the 
following  {\em inversion/reflection equivalence}.
The inversion/reflection equivalence implies
that existence of the unitary operator $U_{\cal I}$ or $U_{\cal R}$
\begin{eqnarray*}
H(x_{\cal I})= U_{\cal I}  ^\dagger  H(x) U_{\cal I},\quad{\rm or }\quad
H(x_{\cal R})= U_{\cal R}  ^\dagger  H(x) U_{\cal R}
\end{eqnarray*}
where $H(x)$ is a complex or a quaternionic hamiltonian
for $x\in S^1$ or $x\in S^3$ respectively.
The inversion in the parameter space is defined as 
$x_{\cal I}=-x$ 
and the reflection is one of the following three, 
$x_{\cal R}=(-x_1,x_2,x_3)$,
$x_{\cal R}=(x_1,-x_2,x_3)$,
and 
$x_{\cal R}=(x_1,x_2,-x_3)$.
As for the $x\in S^1$ case, the reflection is the same as the inversion. 
This is a {\em sufficient} condition for the $\mathbb{Z}_2$-quantization.

Although we use the quaternion notation with the reflection below
(with the KR degeneracy),
it is also true for the inversion and the complex cases.  
The isolated 
KR multiplet, denoted as $\psi(x)$ with the energy $E(x)$, satisfies
$H(x_{\cal R})\psi(x_{\cal R}) =  U_{\cal{R}} ^\dagger H(x)  U_{\cal{R}} 
\psi(x_{\cal R}) =E(x_{\cal R})\psi(x_{\cal R})$
due to the reflection equivalence. 
It implies 
\begin{eqnarray*}
 H(x)  \psi_{\cal{R}} (x) = E(x_{\cal R})\psi_{\cal{R}} (x) 
\end{eqnarray*}
where 
$
\psi_{\cal{R}} (x) = U_{\cal{R}}  \psi(x_{\cal R}) 
$. Since the unitary equivalence between $H(x)$ and $H(x_{\cal R})$ implies 
that all of the eigen values are equal with each other,
we may generically
assume $E(x_{\cal R})=E(x)$ supplementing a unitary transformation of
reshuffling the KR degenerated eigen spaces. 
Now, as for the isolated eigen space of the KR multiplet, 
$\psi(x)$ and $\psi_{\cal{R}}(x)$ are different just in $Sp(1)$ phase, 
which implies that the corresponding Berry connections are 
gauge equivalent,
$\psi_{\cal R}(x)=\psi(x) g$,
 $^\exists g\in Sp(1)$,
\begin{eqnarray*}
\fl\qquad 
a_{\cal R}(x) =\psi_{\cal R} ^\dagger(x) d \psi_{\cal R}(x)
=\psi ^\dagger(x_{\cal R}) d \psi(x_{\cal R}) 
=a(x_{\cal R})= g ^{-1} a(x) g + g ^{-1} d g.
\end{eqnarray*}
Then the generic Berry phases satisfies,
$\gamma_1(A_{\cal R}) \equiv\gamma_1(A)$ and 
$\gamma_3(a_{\cal R}) \equiv\gamma_3(a)$ in modulo 1.
Here note that the $\gamma_1$ and $\gamma _3$  are defined by
the integral over the odd dimensional spaces $S^1$ and $S^3$.
Therefore the generic Berry phases $\gamma _1$ and $\gamma _3$ are 
{\em  odd by the inversion/reflection 
of the parameter space $S^2$ and $S^3$, 
$x\to x_{\cal I}$ 
or
$x\to x_{\cal R}$, 
} as 
$
\gamma_1(A_{\cal I}) =\gamma_1(A_{\cal R}) =
-\gamma_1(A)$ and 
$\gamma_3(a_{\cal I}) =\gamma_3(a_{\cal R}) =
-\gamma_3(a)$. 
Therefore we have a $\mathbb{Z}_2$-quantization of the Berry phases as 
\begin{eqnarray*}
\gamma_1(A_{\cal I}) 
&  \equiv \gamma_1(A_{\cal R}) \equiv
\gamma_1(A)=
0,1/2\quad ({\rm mod}\ 1)
\\
\gamma_3(a_{\cal I}) 
& \equiv 
\gamma_3(a_{\cal R}) \equiv 
\gamma_3(a)=
0,1/2\quad ({\rm mod}\ 1) 
\end{eqnarray*}

\subsection{Chiral symmetry for the minimum models}
\label{sec:chiraMin}
The chiral symmetry of the minimum models discussed before are 
typical example of the systems with 
the inversion equivalence since the anti-commutators
for the unitary $\Gamma=\Gamma ^\dagger$ and $\hc $/$\hq$ are rewritten as
\begin{eqnarray*}
\Gamma ^\dagger \hc (\bi{R})  \Gamma
&=-\hc (\bi{R}) = \hc (-\bi{R}) =  \hc (\bi{R}_{\cal I}), \\
\Gamma ^\dagger \hq (\bi{Q})  \Gamma
&= -\hq (\bi{Q}) = \hq (-\bi{Q}) = \hq (\bi{R}_{\cal I})
\end{eqnarray*}
where the models are defined on the equators as 
$\bi{R}\in S^1$ and $\bi{Q}\in S^3$.
This is what we need for the $\mathbb{Z}_2$-quantization of $\gamma _1$ and
$\gamma _3$. 
We explicitly confirm it by direct calculations below.

Let us first consider a generic case without the KR degeneracy. In the north pole gauge, the multiplet at the equator $R_3=0$, $|z|=1$ is given as
$\Psi_N =\case {1}{\sqrt{2}}\mvec{1}{\bar z}$. Then we have 
$A_N = \case 1 2 z d \bar z = 
\case 1 2 g_{\mathbb{C}} ^{-1} d g_{\mathbb{C}}$, $(g_{\mathbb{C}}=\bar z\in S^1)$.
 It implies $\gamma _1(A_N)= \case 1 2 W_{S^1}(g_{\mathbb{C}})=-1/2$.
If we take the south pole gauge, we have 
$\Psi_S=\case 1 {\sqrt{2}}\mvec{z}{1} $, 
$
A_S=\case 1 2 \bar z d z=-\case 1 2 z d \bar z=-A_N$, $(\bar z z=1)$.
It implies $\gamma _1(A_S)=+ \case 1 2 \equiv \gamma_1(A_N) $, (mod $1$).
 It is consistent with the general consideration and the 
 $\mathbb{Z}_2$-quantization.

With the KR degeneracy, the connection is obtained just by 
replacing $z$ to $q$. 
Then we have the Berry connections in
the two gauges as 
$a_N= \case 1 2  q d \bar q$ and 
$a_S= \case 1 2 \bar q d q$.
Note here that $a_S\ne -a_N$ which is different from the case without
the  KR degeneracy. 
Then using 
 $d q =- q d \bar q q $ ($\bar q q=1$, $d\bar q q =-\bar q dq$ ) and
$da _N= \case 1 2 d q d\bar q 
=-\case 1 2 q d\bar q q \cdot d\bar q 
=-\case 1 2 (q d\bar q )^2$, 
 we have
\begin{eqnarray*}
\fl\qquad 
\omega _3(a _N) &= T(a_N da_N+ \frac {2}{3} a_N^3)=
T\big( -\frac {1}{4} (q d\bar q )^3+ \frac {1}{12}  (q d \bar q )^3\big)
=-\frac {1}{6} T( q d\bar q )^3
\\
\fl\qquad 
\gamma _3(a_N) &= \frac {1}{48\pi^2} \int_{S^3}T (g_{\mathbb{H}} ^{-1} d
g_{\mathbb{H}} )^3 = \frac {1}{2} W_{S^3}(g_{\mathbb{H}})
=-\case 1 2, \quad g_{\mathbb{H}}\in Sp(1).
\end{eqnarray*}
Similarly we have $a_S=\case 1 2 \bar q d q=- \case 1 2 \bar q\cdot 
q d\bar q q=
-\case 1 2 d\bar q  q$,
$da _S=  \case 1 2 d \bar q d q
= 
-\case 1 2  d \bar q \cdot q d \bar q q=- \case 1 2 (d \bar q q)^2$ and 
\begin{eqnarray*}
\fl\qquad 
\omega _3(a _S) &= T(a_Sda_S+ \frac {2}{3} a_S^3)
= T(\frac {1}{4}(d\bar q q)^3- \frac {1}{12} (d \bar q q)^3\big)
=\frac {1}{6} T( d \bar q q)^3=\frac {1}{6} T(q d \bar q )^3
\\
\fl\qquad
 \gamma _3(a _S) &=-\gamma _3(a_N)=
 \case 1 2 \equiv \gamma _3(a_N) \quad{\rm mod}\, 1
\end{eqnarray*}
It again confirms the $\mathbb{Z}_2$-quantization of 
the 
quaternionic 
minimum model
with the chiral symmetry. 

\subsection{Reflection and TR invariant system without  KR degeneracy}
\label{sec:TRref}
The quantization of the $\mathbb{Z}_2$ Berry phase discussed 
in Sec.\ref{sec:tr}
\cite{Hatsugai06a} can be considered as the quantization
 due to the reflection equivalence discussed 
in Sec.\ref{sec:inv} when the parameter introduced is the $U(1)$ twist $e^{i x}$ 
 and the other parameters are all real.
It is simply due to the following observation of the TR invariance
\begin{eqnarray*}
\Theta ^{-1} H(e^{i x}) \Theta = U ^\dagger H(e^{-i x}) U =H(e^{i x})
\end{eqnarray*}
where $U$ is a unitary operator to change
 $c_{i \sigma }\to (-)^{(1-\sigma)/2} c_{i-\sigma }$ for the fermions and 
the spins
 $\bi{S}_i =
 \case 1 2 \bi{c}_i ^\dagger \bi{\sigma } \bi{c}_i$, 
$\bi{c}_i ^\dagger =(c_{i {\uparrow} } ^\dagger ,c_{i {\downarrow} } ^\dagger )$.
This is just the inversion or the reflection equivalence
as discussed in
Sec.\ref{sec:inv}.

\section{Topological Charge and nonlinear $\sigma $-models}
\label{sec:sigma}
Finally in this section, let us   calculate  topological charges
of the nonlinear $\sigma$-model
\cite{Minami80,Wilczek83,Wu84,Wu88,Az91,Demler99}
 as applications
of the gauge invariant forms of the Chern numbers
$C_1$ and $C_2$ in Sec.\ref{sec:minimum}.

\subsection{Topological charge without  KR degeneracy
\cite{Wilczek83,Wu84,wenzee88}}

Let us start by considering a parameter $x$ dependent
two component normalized state
 $\Psi(x) =\mvec{z_1}{z_2}$, $\Psi ^\dagger \Psi=1=
|{\rm Re}\, z_1|^2
+
|{\rm Im}\, z_1|^2
+
|{\rm Re}\, z_2|^2
+
|{\rm Im}\, z_2|^2
$,
which defines $S^3$. 
Then following 3 real quantities $n_1,n_2,n_3$ are defined 
as a $CP^1$ representation of $n_i$, ($i=1,2,3$ )
\begin{eqnarray*} 
\fl\qquad \bi{n}(x) = \mvecthree{n^1}{n^2}{n^3}
= 
\Psi ^\dagger
\mvecthree{ \sigma^1}{ \sigma^2}{ \sigma^3}
\Psi
= 
\mvecthree
 {\Psi ^\dagger \sigma ^1 \Psi }
 {\Psi ^\dagger \sigma ^2 \Psi }
 {\Psi ^\dagger \sigma ^3 \Psi }
= 
\mvecthree
 {\Tr_2 \sigma ^1 P}
 {\Tr_2 \sigma ^2 P}
 {\Tr_2 \sigma ^3 P}
\end{eqnarray*}
where $\sigma _a=\sigma ^a$ and the 
 projection, $P(x)=\Psi\Psi ^\dagger$, into the subspace spanned by $\Psi(x)$
is introduced. 

Since $\Tr P=\Psi ^\dagger \Psi =1$, the projection is expanded 
as
$
P = \frac {1}{2} E_2 +  {P_i} \sigma ^i
$. 
The coefficients are given 
as 
$
P_i = {\rm Tr \,} P \case 1 2 \sigma ^i =  \case 1 2 n^i
$.
Now we have rewritten $P
= \frac {1}{2} (E_2+ n_i \sigma ^i) = \frac {1}{2} \big(E_2+ \hc(\bi{n})\big) 
$
and 
$\hc =\bi{n}\cdot\bi{\sigma }=2P-E_2$.
Then
$\hc^2=4P-4P+E_2=E_2=n_i \sigma _i n_j \sigma _j=n_in_i+\sum_{i<j}n_in_j
\{\sigma _i,\sigma _j\}=|\bi{n}|^2E_2$. It implies
$|\bi{n}|^2=1$.
Therefore the state $\Psi$ can be considered as a positive energy 
eigen state of  $\hc $ by identifying $\bi{n}=\bi{R}$.
It makes a $CP^1$ representation of the $SO(3)$ nonlinear $\sigma $-model.

Using this decomposition of the three  vectors $\bi{n}$, 
let us discuss the  topological charge of the current
\begin{eqnarray*} 
J^\mu =
 \frac {1}{8\pi}  \epsilon^{\mu \nu \lambda } \epsilon_{abc} 
n^a \partial _\nu n^b \partial _\lambda n^c
\end{eqnarray*}
The topological charge is evaluated as 
\begin{eqnarray*} 
\fl\qquad 
Q_{\mathbb{C}}
&=  \int dx^1dx^2 J^3
=   \frac {1}{8\pi}
\int dx^1dx^2 
\epsilon^{3 \nu \lambda } 
\epsilon_{abc} 
n^a \partial _\nu n^b \partial _\lambda n^c
\\
\fl\qquad 
&=    \frac {1}{8\pi}
\int dx^1dx^2 
\epsilon_{abc} 
(n^a \partial _1 n^b \partial _2 n^c
-
n^a \partial _2 n^b \partial _1 n^c)
\\
\fl\qquad 
&=   \frac {1}{8\pi}
\int 
\epsilon_{abc} 
n^a d n^b d n^c
=    \frac {1}{8\pi}
\int 
\epsilon_{abc} 
({\rm Tr \,} \sigma ^a  P)
({\rm Tr \,} \sigma ^b d P)
({\rm Tr \,} \sigma ^c d P)
\end{eqnarray*} 

Since $dP$ is traceless  $2\times 2 $ hermite matrix as 
$
0 = d 1=d {\rm Tr \,} P={\rm Tr \,} dP
$,
 we can expand $dP$ and $P$ as 
$
dP = dP_a \sigma ^a
$, 
$
P=  \frac {1}{2} E_2 + P_a \sigma ^a
$, ($dP_a,P_a\in \mathbb{C}, \ a=1,2,3$).
Now we have
\begin{eqnarray*} 
Q_{\mathbb{C}} &=    \frac {1}{8\pi}
\int 
\epsilon_{abc} 2^3
P_a
dP_b
dP_c
= 
\frac {1}{\pi}\int  \epsilon_{abc} P_a dP_b dP_c
\end{eqnarray*} 
Also note that
$
\Tr  P dP dP 
= 
P_a dP_b dP_c \Tr \sigma ^a\sigma ^b\sigma ^c
=
 P_a dP_b dP_c  i \epsilon^{abd}\, \Tr  \sigma _d \sigma ^c
=2i \epsilon^{abc}   P_a dP_b dP_c
$.
Therefore we finally have
\begin{eqnarray*}
\fl\qquad 
C_1 
=  \frac {i}{2\pi} \int 
 d \omega _1 
=\frac {i}{2\pi } \int \Tr (P dPdP)
=   \frac {i}{2\pi} \int 
(2i)\epsilon_{abc} P_a dP_b dP_c =-Q_{\mathbb{C}}
\end{eqnarray*} 
It gives a direct relation between the first Chern number
and the topological charge of the SO(3) nonlinear $\sigma $-model.

\subsection{Topological charge with   KR degeneracy
\cite{Minami80,Wilczek83,Wu84,Wu88,Az91,Demler99}}

Similarly with the KR degeneracy, 
let us consider a  $x$ dependent
 four component normalized KR pair,
which is described by the two component quaternionic vector
 $\psi(x) =\mvec{\psi_1}{\psi_2}\in \mathbb{H}^2$, $\psi ^\dagger \psi=1
=
(\psi_1^0)^2
+
(\psi_1^1)^2
+
(\psi_1^2)^2
+
(\psi_1^3)^2
+
(\psi_2^0)^2
+
(\psi_2^1)^2
+
(\psi_2^2)^2
+
(\psi_2^3)^2
$,
which defines $S^7$ where 
$\mathbb{H}\ni
\psi_i=
\psi_i^0
+
\psi_i^1\qi
+
\psi_i^2\qj
+
\psi_i^3\qk
$, $\psi_i^a\in \mathbb{R}$, $(a=0,1,2,3, i=1,2)$. 
Then following 5 real quantities $n_1,n_2,n_3,n_4,n_5$ are defined by 
the  $HP^1$ representation as
\begin{eqnarray*} 
\fl \qquad
\bi{n}(x) = \mvecfive{n^1}{n^2}{n^3}{n^4}{n^5}
= 
\frac {1}{2} 
T 
\psi ^\dagger
\mvecfive{ \Sigma^1}{ \Sigma^2}{ \Sigma^3}{ \Sigma^2}{ \Sigma^5}
\psi
= 
\frac {1}{2} 
\mvecfive
 {T\psi ^\dagger \Sigma ^1 \psi }
 {T\psi ^\dagger \Sigma ^2 \psi }
 {T\psi ^\dagger \Sigma ^3 \psi }
 {T\psi ^\dagger \Sigma ^4 \psi }
 {T\psi ^\dagger \Sigma ^5 \psi }
= 
\frac {1}{2} \mvecfive
 {\Tr  T\, \Sigma ^1 p }
 {\Tr  T\, \Sigma ^2 p }
 {\Tr  T\, \Sigma ^3 p }
 {\Tr  T\, \Sigma ^4 p }
 {\Tr  T\, \Sigma ^5 p },
\\
\fl 
\Sigma^1 = \tmat{0}{\qi}{\bar\qi}{0},
\Sigma^2 = \tmat{0}{\qj}{\bar\qj}{0},
\Sigma^3 = \tmat{0}{\qk}{\bar\qk}{0},
\Sigma^4 = \tmat{0}{1}{1}{0}, 
\Sigma^5 = \tmat{1}{0}{0}{-1}
\end{eqnarray*}
where 
$\Sigma^a=\Sigma_a=(\Sigma^a) ^\dagger $, 
$(\Sigma^a)^2=E_2$, $\{\Sigma_a,\Sigma_b\}=0$, 
$\Sigma^a\Sigma^b\Sigma^c\Sigma^d= \epsilon^{abcde} \Sigma_e$ (when
 $a,b,c,d,e$ all different)
and $p(x)=\psi\psi^\dagger$ is a
 projection, 
 into the subspace spanned by the 
KR pair $\psi(x)$.

Since $ \Tr T\, p=T\psi ^\dagger \psi =2$, 
the projection is expanded 
as
$
p = \frac {1}{2} E_2 +  {p_a} \Sigma ^a
$. 
The coefficients are given 
as 
$
p_a = {\rm Tr \,} T\, ( \Sigma ^a p)/4 = n^a/2
$.
Now we have rewritten $p
= \frac {1}{2} (E_2+ n_a \Sigma ^a) =
 \frac {1}{2} \big(E_2+ \hq(\bi{n})\big) 
$
and 
$\hq =\bi{n}\cdot\bi{\Sigma }=2p-E_2$.
Then
$\hq^2=4p-4p+E_2=E_2=n_i \Sigma _i n_j \Sigma _j=n_in_i E_2+\sum_{i<j}n_in_j
\{\Sigma _i,\Sigma _j\}=|\bi{n}|^2E_2$. It implies
$|\bi{n}|^2=1$.
Therefore the state $\psi$ can be considered as a positive energy 
KR multiplet of  $\hq $ by identifying $\bi{n}=\bi{Q}$. 
It establishes the relation for the $HP^1$ representation of the
$SO(5)$ nonlinear $\sigma $-model.

Again using this decomposition of the five  vectors $\bi{n}$, 
let us discuss the  topological charge $Q_{\mathbb{H}}$ following the 
references\cite{Wu88,Az91,Demler99}
\begin{eqnarray*} 
\fl\qquad
J^{\sigma \tau \omega}
&=   N ^{-1}   \epsilon^{\mu \nu \lambda \kappa \rho \sigma \tau \omega}
 \epsilon_{abcde} 
n^a \partial _\mu n^b \partial _\nu n^c
\partial _\lambda   n^d\partial _\rho  n^e
\\
\fl\qquad
 Q_{\mathbb{H}} &=
\int dx^1dx^2dx^3dx^4\, J^{567}
= 
N ^{-1} 
\int  dx^1dx^2dx^3dx^4\, 
\epsilon^{\mu \nu \lambda \rho 567}
 \epsilon_{abcde} 
n^a \partial _\mu n^b \partial _\nu n^c
\partial _\lambda   n^d\partial _\rho  n^e
\\
\fl\qquad
&= 
N ^{-1} \int 
 \epsilon_{abcde} 
n^a d n^b d n^c
d n^d d n^e
\\
\fl\qquad
&=    
N ^{-1} 2^{-5}
\int 
\epsilon_{abcde} 
(\Tr T\,\Sigma ^a  p)
(\Tr T\,\Sigma ^b d p)
(\Tr T\,\Sigma ^c d p)
(\Tr T\,\Sigma ^d d p)
(\Tr T\,\Sigma ^e d p)
\end{eqnarray*} 
where $N$ is  a normalization constant.

Since $dp$ is traceless quaternionic and hermite
$
0 = d 1=d {\rm Tr \,} p={\rm Tr \,} dp
$,
 we can expand $dp$ and $p$ as 
$
dp = dp^a \Sigma_a$, 
$
p=  \frac {1}{2} E_2 + p^a \Sigma_a
$, ($p^a,dp^a\in \mathbb{R}, a=1,\cdots,5$).
Now we have
$
Q_{\mathbb{H}} = 
2^5 N ^{-1} 
\int  \epsilon_{abcde} p^a dp^b dp^c dp^d dp^e
$.
Also we can show
$
\Tr T\, (p dp dp) ^2
=  4  \epsilon_{abcde}   p^a dp^b dp^c dp^d dp^e
$.
Therefore we  have
\begin{eqnarray*}
\fl\qquad
C_2 
&=
 -  \frac {1}{8\pi^2} 
\int \Tr T\, (p dp dp)^2
=  
-\frac {1}{2\pi^2}
\int  \epsilon_{abcde}  p^a dp^b dp^c dp^d dp^e
\propto  Q_{\mathbb{H}}
\end{eqnarray*} 

This is again the direct relation between the second Chern number 
of the $HP^1$ model 
and the topological charge of the $SO(5)$ nonlinear $\sigma$-model.

\section{Acknowledgments}
We thank discussions 
with  T. Fukui, H. Aoki, H. Katsura,  M. Arai,  I. Maruyama, M. Arikawa 
and T. Fujiwara about related materials.
The work is supported in part by Grants-in-Aid for Scientific 
Research, No.20340098, No.20654034 from JSPS and No.220029004 (Physics 
of New Quantum Phases in Super-clean Materials) and 
No.20046002 (Novel States of Matter Induced by Frustration) 
on Priority Areas from MEXT (JAPAN).
The work was also supported in part by the National 
Science Foundation under Grant No. PHY05-51164
at KITP (Santa Barbara) where a part of the work was done during
the mini program ``Quantum Spin Hall Effect and Topological Insulators''.

\section*{References}


\providecommand{\newblock}{}

\end{document}

%% file: mydef.tex
\newcommand{\eas}[0]{\begin{eqnarray*}}
\newcommand{\eae}[0]{\end{eqnarray*}}
\newcommand{\les}[0]{\begin{equation}}
\newcommand{\lee}[0]{\end{equation}}
\newcommand{\leas}[0]{\begin{eqnarray}}
\newcommand{\leae}[0]{\end{eqnarray}}

\newcommand{\tmat}[4]
{\small
\left[
\begin{array}{cc}
#1 & #2 \\
#3 & #4 
\end{array}
\right]
}
\newcommand{\mmat}[4]
{
\left[
\begin{array}{cc}
#1 & #2 \\
#3 & #4 
\end{array}
\right]
}

\newcommand{\mvec}[2]
{
\left[
\begin{array}{c}
#1  \\
#2  
\end{array}
\right]
}

\newcommand{\mvecthree}[3]
{
\left[
\begin{array}{c}
#1  \\
#2  \\
#3  
\end{array}
\right]
}
\newcommand{\mvecfive}[5]
{
\left(
\begin{array}{c}
#1  \\
#2  \\
#3  \\
#4  \\
#5  
\end{array}
\right)
}